# Software Dependability Measurement at the Age of 36

Robert V. Binder [*]

RBSC Corporation

Pittsburgh, Pennsylvania, USA

rvbinder@rbsc.com

## ABSTRACT

Thirty-six years after the first edition of IEEE standard 982.1, Measures of the Software Aspects of Dependability, the third edition focuses on the measurement of in-service software dependability. This article explains how this new point of view evolved and shaped the third edition's guidance for software dependability measurement.

***Keywords*** dependability reliability availability supportability recoverability safety security performance

## 1 Introduction

Software dependability is the reliability, availability, supportability, and recoverability of a software system. Understanding what this means, how to characterize it, and how to achieve it has been of perennial interest to software and system engineers, system sponsors, system users, and researchers. For 36 years, IEEE Standard 982.1 has provided definitions and formulas for dependability measurement and improvement. The first edition appeared in 1988; the second in 2005. This article discusses the forthcoming 2024 edition and its significant changes.

Contemporary software engineers may note that dependability measurements are well-defined, and development practices once advocated (or decried) for dependability evolved when multi-core processors were found only on supercomputers, OS virtualization was dead, object-oriented programming was at the peak of its hype cycle, AI had entered yet another winter, and agile was for gymnasts.

---

[*] Mr. Binder served as the Chair of the IEEE Working Group that produced 982-2024.

SOFTWARE DEPENDABILITY MEASUREMENT AT THE AGE OF 36

Although software engineering practices and technologies have changed greatly since the 1988 edition, software dependability remains the sine qua non of quality attributes. The business, safety, security, and societal risks of undependable software have increased apace with its viral ability to find and exploit new application niches. Despite technology upheavals (and in many cases because of them), the need to know how much we can depend on software-intensive systems is greater than ever. Regardless of development strategies, enabling technology stacks, and the human purpose of application systems, undependable software is trouble for which no one wishes and of which many cannot abide.

Dependability measurement should provide consistent and credible answers to questions like "Is system $x$ more (or less) dependable than system $y$? How so?" or "Is my system's dependability improving or deteriorating over successive versions? How so?"

The new edition of 982 is therefore focused on the black-box measurement of in-service software dependability and not on prescriptions for development or operational practices. It introduces, among other things, criteria for on-the-fly anomaly scoring so that emergent misbehavior, safety anomalies, and security exploits are just as visible as hard crashes. It is technology and application agnostic, intended for any kind of software system, including those using technologies identified as "AI".

The 1988 first edition of 982.1 collected then-novel development-oriented terms and formulas into a concise glossary and provided a companion guide to development practices in 982.2 (this guide was withdrawn in 2002.) Seventeen years on, the 2005 second edition presented revised formulas derived from leading-edge research and practice.

The forthcoming third edition appears 19 years after the second edition. It is retitled as the IEEE Standard for Measures of the Software Aspects of Dependability and renumbered as IEEE Standard 982.[†] As many other sources, notably IEEE Standard 1633-2016 (*IEEE Recommended Practice on Software Reliability*), provide comprehensive and current guidance for software dependability practices during development, the third edition focuses on the black-box measurement of in-service dependability. While this edition carries forward some of the 2005 edition's formulae, it sharpens its focus, makes assumptions explicit, and provides a novel framework for consistent and actionable measurement. It introduces the In-service Reference Model (ISRM), a support action taxonomy, and criteria for on-the-fly anomaly scoring. It structures any class of anomaly or failure (including safety and security) so that they can be characterized with established reliability formulae. It calls for explicit dependability requirements with thresholds quantified as percentiles or confidence intervals, not simply as averages. It leaves the understanding of failure inherited from the realities of physical systems and provides a new formulation that accurately reflects software-specific failure phenomena. It recognizes the critical distinction between software support and software maintenance, setting a bright line between downtime arising from in-service actions and work that must be done at upstream facilities.

This new point of view arose as a result of the 982.1 Working Group's research and discussions about how to make software dependability measurement useful for the pervasive and critical role of all kinds of software in 21st-century life. This article explains how it evolved and shaped the third edition's guidance for software dependability measurement.

---

[†] The third edition is referred to as 982-2024 in this article. As this article went to press, the IEEE had not formally assigned a number to the third edition and released it for public use, so the standard's numerical designator may change when it is published.

# 2 What's New in Software Dependability?

## 2.1 Compliance criteria

982-2024 uses normative statements with the imperatives *shall*, *should*, and *can* to designate artifacts and actions to be addressed for claiming self-compliance with this standard. The purpose of this normative tagging is to give specific meaning to claims like "Our system meets the dependability criteria of IEEE Standard 982-2024."

In this article, normative artifacts or actions are said to be "called for" or that the standard "calls for" such.

## 2.2 In-service, black-box focus

982.1-2005 included predictive software reliability growth models primarily as development tools. Some of them relied on white-box measurements such as source code defect counts. As 982-2024 is focused on measurement of in-service behavior, it does not include predictive models and does not call for any kind of white-box measurements. In part, this reflects the fact that comprehensive guidance for use of predictive software reliability growth models can be found in IEEE Standard 1633-2016.

## 2.3 A harmonized definition

982.1-2005 defined dependability as the "Trustworthiness of a computer system such that reliance can be justifiably placed on the service it delivers. Reliability, availability, and maintainability are aspects of dependability." However, as explained in the following, trustworthiness is not in the scope of 982-2024. Consistent with the third edition's black-box and in-service perspective, supportability has replaced maintainability. Reliability and availability are understood as quality attribute measurements, not aspects. As fault tolerance can significantly improve software dependability, 982-2024 now includes recoverability to measure its effect. 982-2024 now defines software dependability as "the extent to which measured values of reliability, availability, recoverability, and supportability meet or exceed their required thresholds during an in-service interval of a software system of interest."

982-2024 supports measurement of these quality characteristics with definitions of pertinent variables and formulae listed at the end of this article: 11 for reliability, 19 for availability, 9 for supportability, and 8 for recoverability. All of these items are expressed in terms of the modes and events of the ISRM and a common timing model.

This understanding is consistent with the International Electrotechnical Commission's (IEC) most recent definition of dependability for hardware/software systems.[1] The IEC defines dependability of an "item" as its "ability to perform as and when required...dependability includes availability, reliability, recoverability, maintainability, and maintenance support performance, and, in some cases, other characteristics such as durability, safety and security…dependability is used as a collective term for the time-related quality characteristics of an item." Table 1 shows how each element of the IEC definition aligns with 982-2024.

| Element of IEC Dependability | 982-2024 Dependability Model |
|---|---|
| Ability to perform as and when required | All dependability measurements must be traceable to one or more dependability requirements. Traceability from individual capability requirements to dependability requirements must be established. "As" compliance is measured with the failure intensities of each reliability class. "When" is measured with on-demand readiness (availability) and the failure intensity of the performance reliability class. |
| Availability | An in-service interval consists of uptime and downtime. Software availability is the uptime fraction of this interval, expressed as a percent or probability. Downtime is characterized by the duration and type of support actions and the duration of nonsupport inoperability. |
| Reliability | Software reliability characterizes the number, pattern, and frequency of software failures observed during system uptime and of data failures at any time. It is expressed as the failure intensities of the reliability classes safety, security, functionality, performance, and utilization; user-defined classes may be added. |
| Recoverability | Software recoverability characterizes the frequency and extent to which run time impairments are mitigated with built-in fault tolerance capabilities. |
| Maintainability and maintenance support performance | Software supportability characterizes the duration and frequency of in-service support actions. This corresponds to the IEC "maintenance," which refers to the field service of physical systems. In contrast, software maintenance entails upstream debugging and programming that can only be done in a development environment. Software maintainability is not in the scope of 982-2024. |
| Durability | Data durability is measured as the recoverability from data impairments and the extent of data failure lossage. |
| Safety | Software safety is a defined reliability class, traceable to safety requirements. It is measured as safety failure intensity. |
| Security | Software security is a defined reliability class, traceable to security requirements. It is measured as security failure intensity. |
| Used as a collective term | Software dependability is a composite quality characteristic consisting of reliability, availability, supportability, and recoverability measurements. It is not expressed as a single number. |
| Time-related quality characteristics | All software dependability measurements are defined with respect to their time of occurrence within the duration of an interval of interest following an explicit timing model. |

*Table 1 Dependability as defined in IEC and 982*

## 2.4   Why isn't trustworthiness part of dependability?

The measurement of software trustworthiness is not in the scope of 982-2024. While the present-day understanding of trustworthiness includes all the elements of dependability that 982-2024 defines, it also hinges on evaluation of sociopsychological phenomena related to the human perception of software behavior.[2] 982-2024 is focused on the measurement and characterization of in-service software behavior and operation, regardless of how it is perceived. It does not address the measurement of sociopsychological phenomena.

## 2.5 Can dependability be usefully represented with a single number?

982-2024 does not define a formula or algorithm to aggregate its component measurements as a single number. The working group reviewed research and proposals to produce a composite value. No candidate approach (1) had a similarly unambiguous frequency-of-occurrence interpretation as the component measurements, (2) had a comparable grounding in well-established statistical analysis, (3) could plausibly serve as a meaningful and consistent indicator of favorable or unfavorable outcomes, and (4) offered credible evidence of or a plausible case for successful use.

## 2.6 Why was maintainability replaced with supportability?

As software maintainability pertains to upstream (development) actions, it is not within the scope of 982-2024. Downstream (in-service) support actions that result in downtime are in scope. Therefore, 982-2024 uses *supportability* to characterize the effect of support actions on uptime. Software maintainability is no longer included in its definition of dependability.

The concept of software maintainability has its origin in availability and reliability models of physical systems. Maintainability of physical systems reflects the downtime needed to keep them in service (replenishment of consumables, re-calibration, cleaning, etc.) These concepts were adapted for software around 40 years ago to characterize the ongoing work to debug, update, and re-release software. In the software context, maintainability was then associated with the relative cost or difficulty of this work. A good deal of research and practice has been devoted to understanding its contributing factors such as programming style, modularity, documentation quality, and technical debt (to name a few), in the hope of reducing software maintenance. In some cases this understanding included downstream operational support actions such as making data backups, managing licenses, editing configuration data, installing new versions, etc. But, for the most part, software maintainability is concerned with work controlled and performed in development environments. For software, maintainability characterizes upstream effort, not in-service downtime: it is "…the planning for and maintenance of software products or services, whether performed internally or externally to an organization. *It is not intended to apply to the operation of the software*."[3] [Italics added.]

Although maintainability is correlated with in-service availability and reliability, upstream work typically does not result in downtime until an update is provided to an end-use environment. The in-service downtime needed to process an update is often significantly less than the duration of the upstream work to produce, test, package, and distribute that update. As 982-2024 is focused on in-service effects, it is limited to measurement of support actions that have a direct effect on the duration of in-service downtime. Therefore, it does not address the extent or efficiency of upstream work to debug and modify software.

## 2.7 Mean Time Between Software Failures (MTBSF)

Hardware failures result from design errors, manufacturing variation, wear out, deterioration, or damage. Software faults (bugs and vulnerabilities) result from development errors. A software failure results when a fault is subsequently executed with a data state sufficient to trigger an anomalous response. System failures result when either hardware or software (or both) fail.

The hardware understanding of failure is assumed in the statistics mean time to failure (MTTF) and mean time between failures (MTBF). MTTF characterizes failures in a population of non-repairable items, which each cease to be usable upon failing and cannot be repaired. MTBF characterizes failures in a population of repairable items, which each cease to be usable upon failing, but can be returned to use after a repair.

Unlike physical failures that occur once and then can or cannot be repaired, software failures cannot be repaired by a support action, such as restarting after a failure, as unchanged code still contains faults, even if it can still be used. Further, many software failures do not result in the immediate uncommanded cessation of execution, that is, a terminal software failure. Instead, software often continues running after a failure occurs. In contrast, a failed physical item must be replaced or repaired before it can be used again.

The rare exceptions to these basic failure cycles do not obviate a common problem. Persons whose understanding of MTTF and MTBF is rooted in physical system reliability may assume that software failures and their statistics have the same interpretation and implications as physical failures and their statistics. This is comparable to problems caused when quantities of customary units of measure are interpreted as SI units.

982-2024 therefore defines a software-specific understanding of consecutive failures and the statistic mean time between software failures (MTBSF) as one of its measures. MTBSF is the average of the times between the consecutive occurrences of software failures during a time span of interest, regardless of the immediate effect of any such failure. When software failures are assigned to a reliability class, MTBSF can be computed for safety, security, functionality, performance, or utilization. When an MTBSF value is used as a requirement threshold, the standard calls for the specification of its confidence level and/or occurrence percentile.

MTBSF accounts for software failure recurrence and deferred repair that hardware time-to-failure definitions do not allow. The use of MTTF and MTBF is deprecated for software measurements. They remain valid for physical systems provided that MTTF is limited to nonrepairable systems and MTBF is limited to repairable systems.[4]

# 3   No Characterization without Requirements

## 3.1   *What is the role of requirements in dependability?*

982-2024 calls for dependability requirements as a prerequisite for dependability measurement. The significance and utility of dependability measurements is a direct result of how well they reflect the purposes and consequential effects of system operation. Explicit requirements express these understandings and thereby provide a necessary foundation for the collection and interpretation of dependability measurements.

982-2024 notes four kinds of requirements:

1) *outcome*: the description of indirect effects or desired conditions that achieve a system's purpose, for example key performance indicators,
2) *capability*: the specification of particular responses, behaviors, and operating conditions that contribute to outcomes,
3) *revealed*: the recognition of unanticipated behaviors and defining them as acceptable or not,

4) *dependability*: the bounded specification of reliability, availability, supportability, and recoverability thresholds, for example "The system shall achieve a failure intensity of 1 or fewer countable failures for every 10000 mission hours" or "The system shall have no more than 5 minutes of unplanned downtime for every 8760 hours of operation in a supported environment."

982-2024 calls for the development, documentation, and validation of outcome, capability, and dependability requirements as a routine development activity, and for producing revealed requirements from anomalies and bug reports. It calls for bounded thresholds in dependability requirements. It does not otherwise prescribe the form of expression or development process of requirements. This is the only exception to 982's exclusion of development-related topics and artifacts.

## 3.2 What is a revealed requirement and why is it needed?

All extant definitions of reliability (including 982.1-2005) call for pre-specification to distinguish acceptable from unacceptable system behavior. For example, the SQuaRE standards define reliability as the "degree to which a system, product or component performs *specified* functions under *specified* conditions for a *specified* period of time."[5] The noted IEC definition defines dependability as the "ability to perform *as and when required*." [Italics added.]

So what are we to make of behavior that is clearly undesirable, but which no capability requirement explicitly prohibits or requires? Or of a claim that a system is "working as designed," when it has a significantly detrimental effects? Examples abound: an aircraft's navigation system crashed as it flew west through the international date line, causing all other mission software to fail[6]; a test data generator for an algorithmic trading system was installed in production instead of the module that responded to live market conditions, causing a $440 million loss in a few hours[7]; a robo-taxi was totaled when it did not yield the right of way to a fire truck with lights flashing and sirens howling.[8]

Although it is common practice and common sense to recognize detrimental behaviors as failures even when they are not explicitly prohibited, all extant definitions of reliability include the qualifier "required" or "specified." This plainly means that absent an explicit requirement to the contrary, a detrimental behavior cannot be considered a failure. To close this loophole, 982-2024 introduces the concept of a *revealed requirement*.

Revealed requirements have equal standing with explicit requirements, closing the prespecification loop-hole. They permit recognition of unspecified detrimental behavior (or the detrimental absence of behavior) as revealing a requirement that would have prohibited that behavior. Such recognition is often grounded in expectations expressed in outcome requirements, as well as in unanticipated special cases of explicit capability requirements. Unspecified detrimental behavior can thus be counted as a failure even if not prohibited by preexisting requirements, or when existing requirements have not been revised to reflect system changes.

Revealed requirements should be used as a tool to close a gap that occurs all too often in practice. They should not be taken as a reason to diminish, in any way, the necessity of competent good faith engineering to produce, validate, and maintain comprehensive capability requirements in the first place.

## 3.3 No more mean surprises

Reliability and availability requirements are often specified as mean-time-to-*x*, for example MTBSF. When used as a statistic of observed behavior, interpretation is straightforward–they are simply an arithmetic average of the times of certain events. However, when used as an unbounded threshold value in a requirement statement, they can obscure the fact that detrimental outliers are implicitly acceptable.

When the times between consecutive failure occurrences are normally distributed, half will be unfavorable, i.e., *below* the threshold. If they are exponentially distributed (as is often the case with failure data), about two-thirds will fall below the threshold.

For example, a reliability requirement states that "The guidance system shall have a mean time between software failures of 220 operational hours, per mission." In-service data yields an MTSBF of 230 hours. However, suppose that 17% of the failures occurred within less than an hour of each other, resulting in a *median* TSBF of 96.0. While the producer of this system can claim it has met its reliability threshold requirement, a customer or user would very likely complain of poor reliability.

982-2024 therefore calls for reliability requirements to specify either a confidence interval or percentile for threshold values, not simply a mean value. A percentile threshold specifies the acceptable fraction of failure occurrence. For example, "The guidance system shall have a mean time between software failures of 100 operational hours at the 95th percentile." In the example, the lower bound of the 95% confidence interval for an MTBSF of 220 would be about 100 h. In other words, the MTBSF is required to be more than 100 h for at least 95% of the missions; under 100 h is acceptable only if it occurs in no more than 5% of the missions.

Dependability requirements with bounded thresholds make down-side risks explicit for system developers, operators, and users.

## 4   A Model-based Standard

982-2024 uses several models to provide an unambiguous abstract framework of the elements necessary to accurately and consistently make dependability measurements. The first step in making the dependability measurements defined in 982-2024 is to establish system-specific data items for the ISRM.

### *4.1   What is the In-Service Reference Model?*

When is a software system considered to be running? What does "running" mean for present-day system architectures? Was a system stopped as result of a commanded shut-down, a bug in it, a bug or impairment in its run time stack(s), or damage to the stack's physical environment?

Did the system fail and crash or keep running? How many times? To what effect? When can it be considered ready to run, but not running? When does it stop? How do we account for the time when a system is offline because it is not in use, or because it cannot be used? Should a facilities outage or hardware failure that prevents software from doing its job also count as a software failure?

When does software support start and end? What kinds of activities are included in support? Should we count total elapsed time or just direct effort as downtime? Should delays caused by a third party be included? How can we track and analyze support by purpose?

What is the relationship between support, maintenance, and development? What happens when a system is uninstalled? Before it is installed? Can a system fail after it is uninstalled?

While straightforward, these and similar questions cannot be given specific answers without reference to a particular system, its environment, and its requirements. The standard is intended to provide consistent and meaningful dependability measurements, regardless of any measured system's implementation technology and application domain. All of 982-2024's variables and formulas are therefore defined in terms of the abstract operational modes and events of its ISRM.

The ISRM reflects broad experience with many kinds of software technologies and applications, and is intended to support any of them. The standard calls for mapping the pertinent details of a particular system into the ISRM, so that system-of-interest behavior can be consistently tracked and evaluated using 982's dependability formulae.

'The ISRM is expressed as a state chart in *Figure 1*. State chart models represent sequentially constrained and concurrent behavior as hierarchies of states with some graphical shortcuts to simplify presentation. A mode (rounded rectangle) represents a generic operational activity or condition. Concurrent modes are represented when two or more modes at the same level are separated with a dashed line. An event (arrow) represents a system behavior, a support action, or other occurrence that results in a different mode or is processed without a change in mode. In this chart, transitions for some pairs of modes that can switch between each other are represented with a single, double-lined arrow.'

'Mode names are indicated in uppercase: READY, BLOCKED, RUNNING, etc. Events names are indicated with capitalized italics: *Normal Run Response (NRR)*, *Terminal Failure (TF)*, etc. An up-time interval begins when the READY mode starts. It ends when a *Normal Run Termination (NRT), Terminal Failure (TF), Stack Failure (SF)*, or *Operational Failure (OF)* event ends the RUNNING mode. If an SOI can be restarted without any support action, the READY mode results and the uptime interval continues.'

'The ISRM distinguishes the layer that causes a failure: the software system of interest, its runtime stack, or its operational environment. *Non-Terminal Failures (NTF), Terminal Failures (TF),* and *Data Failures (DF)* originate in the SOI layer. A *Stack Failure (SF)* originates in the SOI's runtime software/hardware stack. An *Operational Failure (OF)* originates in the facility or device that supports the runtime stack.'

'If an SOI cannot return to the READY mode because a support action must be performed, it enters and remains in the INOPERABLE mode until a support action event *Support Start (SS)* occurs. The SUPPORTING mode then continues until a *Support Finish (SF)* event occurs, after which the READY mode starts.'

'The ISRM represents anomalous events that can occur even when an SOI is not executing. A *Data Failure (DF)* event represents an occurrence of unpermitted access, data loss, or data corruption that an SOI's configuration is expected to have prevented, even if the SOI it is not executing. The REMOVED mode allows tracking detrimental effects than can arise from the residue of uninstalled software. This residue can be an unexpected source of security failures.'

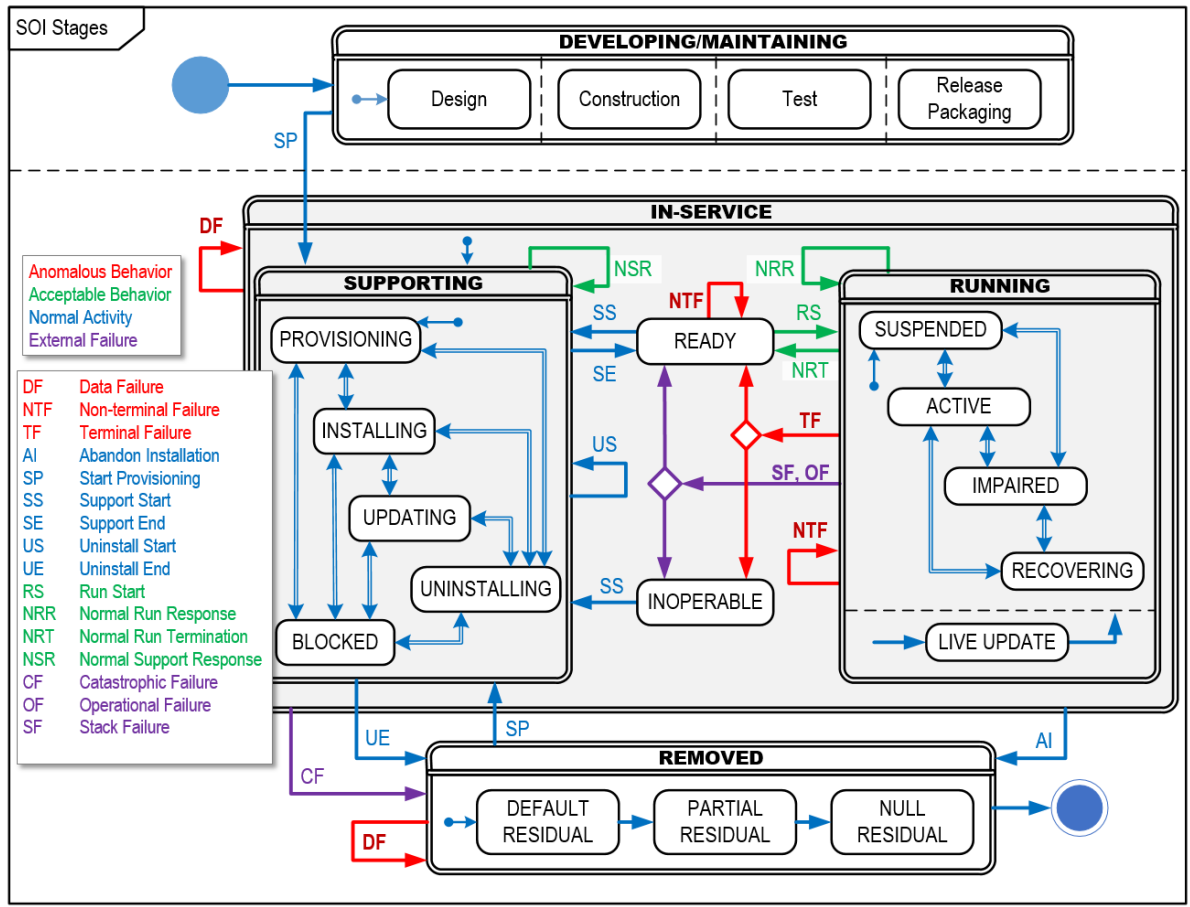

*Figure 1 In-service Reference Model*

## 4.2 The Reliability Class Model

The catch-all failure approach used in 982.1-2005 reports a single reliability measurement for all sufficiently severe failures. This poses four problems. First, it does not distinguish among safety, security, performance, utilization, functional, and non-terminal failures. Second, there are often orders-of-magnitude differences among failure intensity thresholds of each reliability class. For example, suppose no more than 1 in a million missed deadlines in a hard real-time time system is acceptable, but a much lower rate of incorrect image identification is acceptable, 1 in 10,000. These requirements cannot be meaningfully represented with an undifferentiated threshold. Third, severity criteria can be distinctly different. For example, a near-miss is a severity level 2 safety failure, while the severity of excessive utilization of a data center server depends on the imputed value of abandoned transactions. Fourth, classes often have distinct operational considerations for data collection. These differences must be recognized and accommodated to achieve timely, accurate, and consistent collection of class-specific failure data.

The reliability class model explicitly distinguishes the criteria, measurement, and evaluation of canonical quality attributes so that they can be modeled and reported using well-established formulae of reliability analysis. Failure severity criteria, failure intensity reporting, and reliability thresholds can be set and tracked for each reliability class. This framework for specification, monitoring, and remediation makes essential differences visible using a common approach.

The reliability class model consists of $n$ classes, denoted as upper case lambda $\Lambda_n$. The universal reliability class $\Lambda_0$ is the default and subsumes all other classes. Each canonical quality attribute has an explicit class: safety ($\Lambda_1$), security ($\Lambda_2$), functionality ($\Lambda_3$), performance ($\Lambda_4$), and utilization ($\Lambda_5$). The model is extensible; users may add any user-defined reliability class or classes of interest $\Lambda_n$, $n > 5$. Class $\Lambda_0$ also provides downward compatibility for the undifferentiated data sets assumed in 982.1-2005 and typically used in reliability literature.

For each reliability class applicable to a system of interest, the standard calls for its specification comprising 1) an observation strategy, 2) the behaviors to be monitored, 3) classification process and criteria, including severity level cut-off, and 4) the reliability threshold requirements.

## 4.3 How is the Reliability Class Model Used?

The chart in Figure 2a sketches the process for producing a data set consisting of countable failures for each defined reliability class. The observation step on the left-hand side represents an instance of the system of interest that produces observable behaviors while in the READY, RUNNING, or REMOVED modes. It is crucial to record the time each behavior occurs and the duration of each mode. (This chart does not show other operational activities that would typically occur in conjunction with the noted steps, such as system administration, defect reporting, or incident management.)

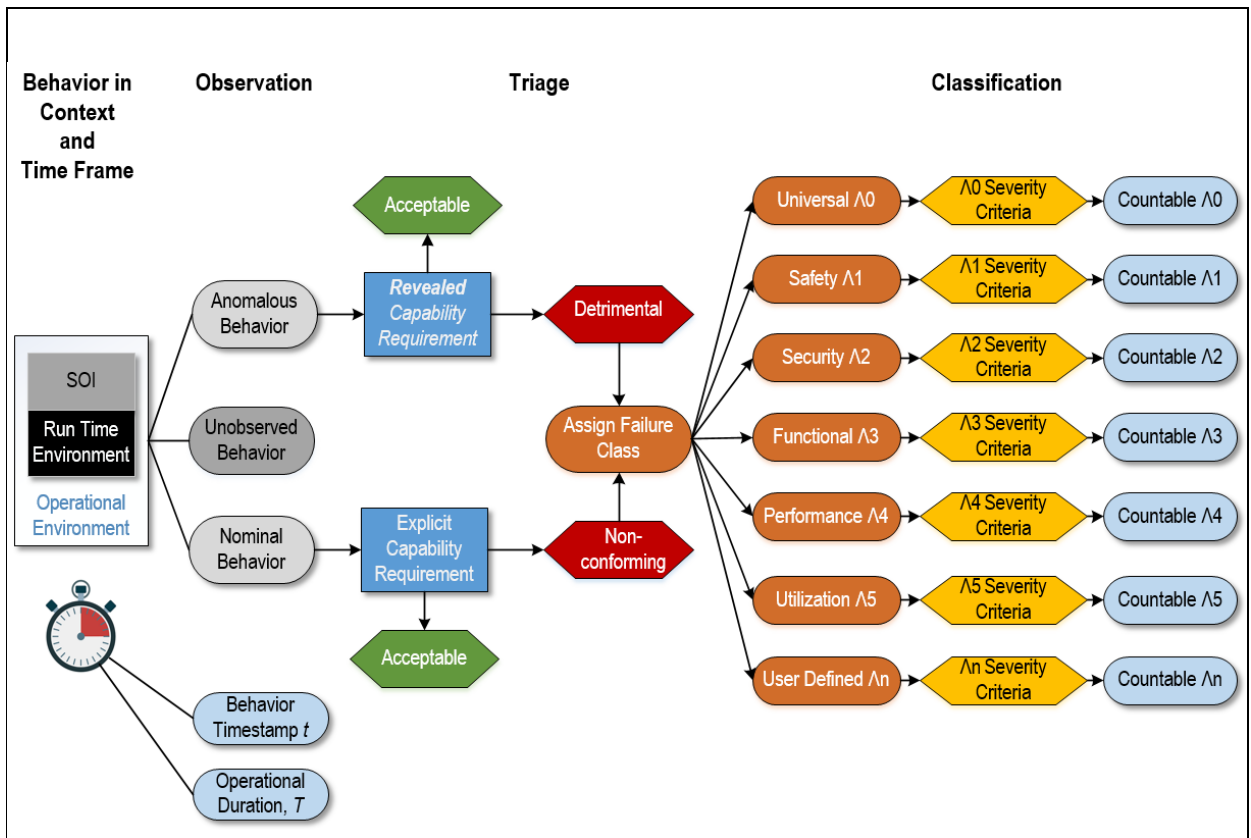

*Figure 2a Observation, Triage, and Classification (continued in Fig 2b)*

In the triage step, observed behavior is categorized as anomalous (not a self-evident failure but suspicious) or nominal (not a self-evident failure and not suspicious). Anomalous behavior may then be evaluated to determine if it is benign or detrimental, in either case producing a revealed requirement as a side-effect. A detrimental anomaly is a failure. Similarly, a nominal (required) behavior may be evaluated as acceptable or unacceptable, often depending on external factors. Unacceptable nominal behavior is considered a failure. The ISRM shows the post-triage failure result as an *NTF, TF, EF,* or *DF*.

Classification is the next step. Failures are assigned to a reliability class. As a particular behavior can violate more than one requirement, a failure can be counted in multiple classes. Next, severity is assessed. 982-2024 defines four generic severity levels: extreme detriment, high detriment, moderate detriment, and minimal detriment (users may substitute any similar ranking criteria.) Following 982.1-2005 and long standing practice, each reliability class must include a severity cut-off (e.g., only extreme and high detriment failures are counted.) If no severity cut-off is set, all failures of a class are counted. If a failure is of sufficient severity, it becomes a *countable failure* of reliability class $\Lambda_n$. The result is a set of countable failure events occurring in period $T$, in each defined reliability class.

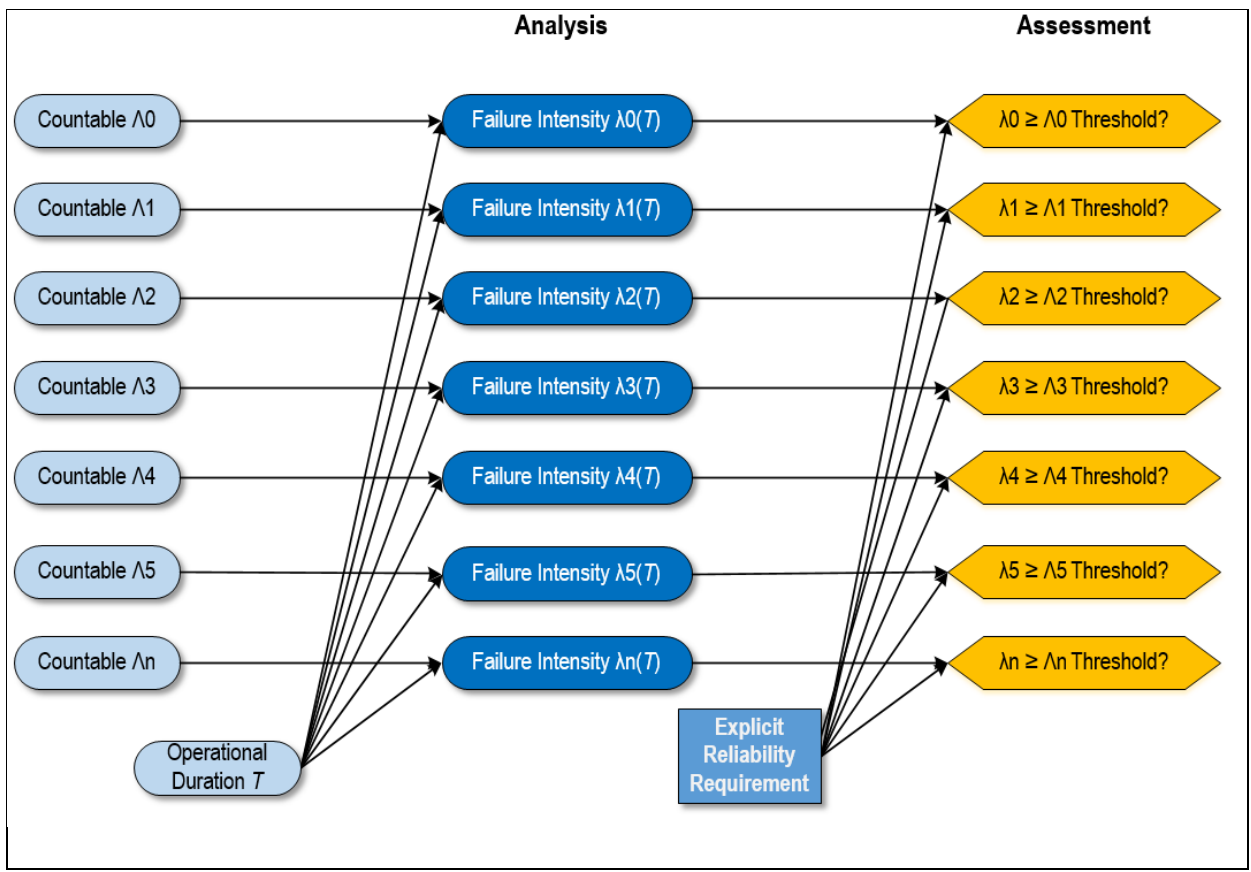

*Figure 2b Analysis and Assessment*

The chart in Figure 2b continues with analysis and assessment of these countable failures. In the Analysis step, the essential reliability measurement of failure intensity $\lambda_n$ is produced. This is simply the occurrence rate of class $\Lambda_n$ failures during period $T$.

In the final step, (assessment), the question "*Is the observed failure intensity $\lambda_n$ lower than the required threshold rate $\Lambda_n$ (a favorable outcome), or does it exceed the threshold (unfavorable)?*" can be answered. To support this determination, 982-2024 calls for explicit dependability requirement statements that express the minimum acceptable level of failure intensity with a boundary, by reliability class. This is a change from 982.1-2005, which was entirely silent about dependability requirements.

### *4.4 How 982-2024 addresses safety and security*

982-2024 explicitly accounts for safety and security as components of dependability. This brings a uniform approach to specifying and monitoring failure intensity thresholds for safety- and security-related capabilities, instead of spawning separate and esoteric realms of development, monitoring, and mitigation.

- First, the reliability class model includes security and safety classes. As with all other reliability classes, this calls for explicit dependability requirements with bounded thresholds for security and safety.
- Second, safety or security failure intensity (that is, safety or security reliability) is visible both individually and as part of a system's dependability constellation. These values can be tracked over time and readily compared with those of other systems, as well as an SOI's other reliability classes.
- Third, the 982-2024 concept of a revealed requirement allows the on-the-fly identification and mitigation of security and safety failures to be integrated with security and safety engineering. For security, revealed requirements are particularly well suited to the inherently opportunistic nature of malicious and covert exploits, such as zero-day attacks and advanced persistent threats. For safety, revealed requirements are well suited to the established safety practice of tracking near misses as leading indicators of safety incidents.
- Fourth, class-specific severity cut-offs establish explicit criteria for prioritizing failure resolution, avoiding the inefficiency of ad hoc case-by-case adjudication.
- Fifth, the ISRM recognizes that the post-operational residue of an uninstalled or decommissioned system is often the target of security exploits. It therefore includes a REMOVED mode in which exploits that rely on a residue (*Data Failures*) can occur. This can support monitoring and tracking of exploits that prey on data residues, a type of vulnerability that is often given short shrift.

## 5 The Structure of Downtime

The ISRM shows how an in-service interval is partitioned into downtime and uptime. Uptime is the duration of the READY and RUNNING modes. Downtime is everything else, namely the modes INOPERABLE, SUPPORTING, and BLOCKED. The relationship between these modes and the kinds of activities they can involve is shown in Figure *3*.

The INOPERABLE mode represents a system that cannot be restarted unless some support action is taken. The SUPPORTING mode represents support work in progress. This work may have been planned or unplanned. For example, support actions to establish a failure workaround are typically unplanned; making a routine backup is typically planned.

Support actions can be tagged according to purpose: adaptive, additive, corrective, emergency, perfective, or preventative. These categories are useful for planning and managing a support process. Retrospective analysis of the proportions and frequency can provide insights useful for process improvement.

Support work may be blocked for a number of reasons. BLOCKED mode is useful for accounting for delays that extend downtime. Passing along or absorbing the costs of the delays is often a contractual obligation of a system's provider, its sponsor, or its operator. The nature of a delay can therefore be significant.

'An administrative blockage is a result of a procedural action that interrupts a support action. For example, a stop work order is issued owing to a payment dispute. A technical blockage results when a blocking bug in a system of interest prevents completion of a system of interest support action. A logistic blockage results when resources necessary for a support action are unavailable. An external blockage results when events outside the control of the developer and/or operator disrupt or prevent completion of support actions.'

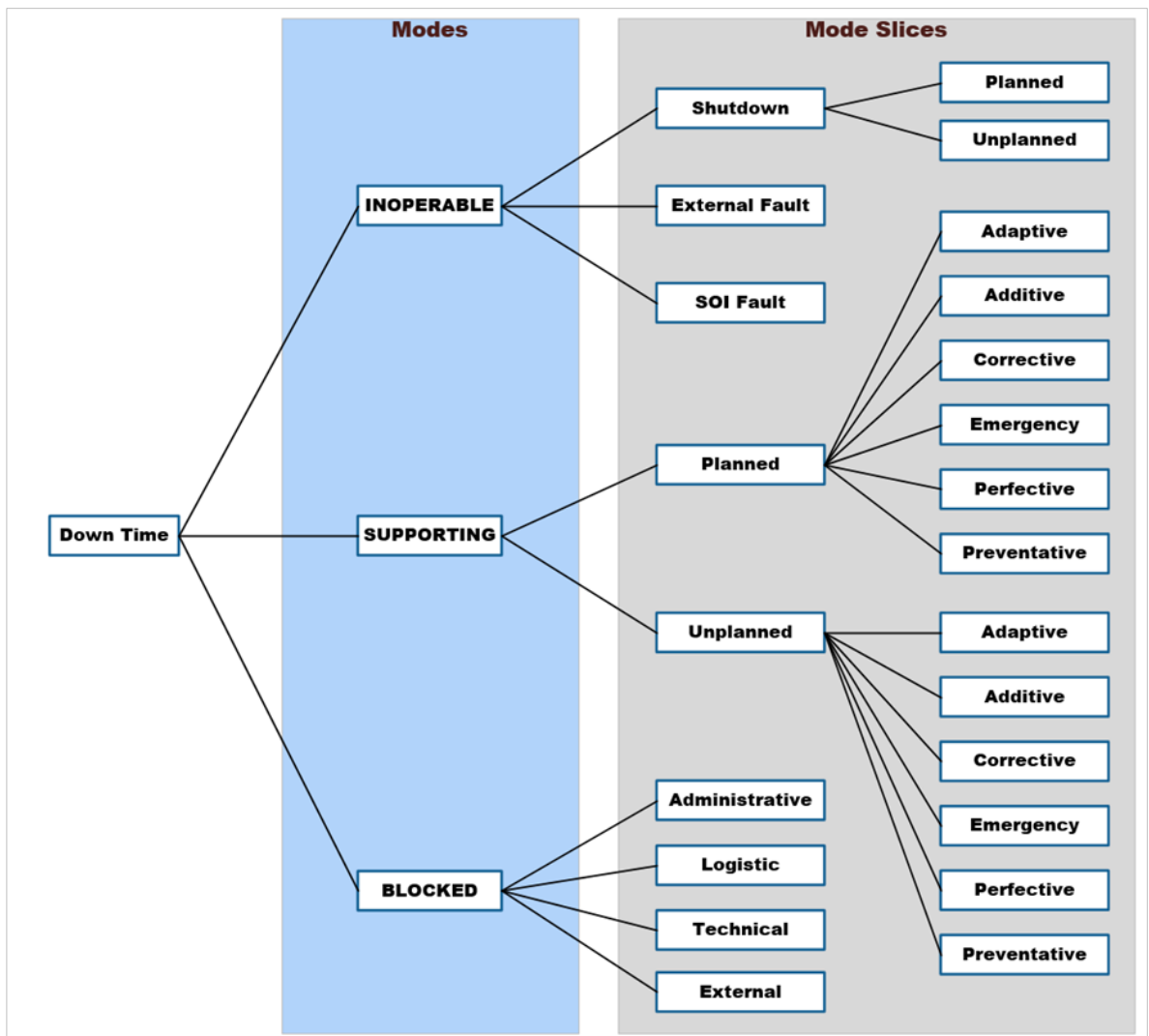

*Figure 3 Downtime Modes and Slices*

There are sixteen possible combinations of schedule, purpose, and blockage slices. Each defines a distinct downtime subset. Table 2 shows six notional examples.

| Support Action | Schedule? | Purpose? | Blockage? |
|---|---|---|---|
| Install bug-fix release and vulnerability patches during planned downtime. | Planned | Corrective | None |
| Change user interface to operate in new locale during planned downtime. | Planned | Adaptive | None |
| Install new feature release during planned downtime. | Planned | Perfective | None |
| Perform emergency rollback to previous version when a new version fails. | Unplanned | Corrective | None |
| Reconfigure to accommodate security hotfix. | Unplanned | Adaptive | None |
| Bug in third-party configuration script; cannot install new version. | Planned | Perfective | Technical |
| Host data center destroyed by hurricane; cannot perform rollback until standby data center is online. | Unplanned | Corrective | External |

*Table 2 Example Support Actions and Slices*

# 6   What's in the Standard?

The standard includes concise definitions of 75 terms, guidance for compliance self-assessment, informative guidance about supporting topics, and normative definitions for the formulas and variables of each dependability component.

The supporting topics are 1) abstract data model, 2) criteria for defining the system of interest, 3) outcome, capability, revealed, and dependability requirements, 4) the in-service reference model, 5) the reliability class model, 6) the downtime support slice model, and 7) the timing model.

Each dependability component consists of formulas and variables bounded by a time span of observation, $T$. The timing model specifies how to delimit the start and stop of any span of interest.

## 6.1   Reliability

'Software Reliability is the extent to which a system of interest has operated without countable failures in a specified environment for a specified time span. It is a quantitative component of dependability expressed as a requirement threshold, as a probability (prediction), or as a measurement obtained during test or in-service operation (observation).'

'Observed reliability is expressed as failure intensity, the ratio of the number of failures observed in a span to its duration.'

'In terms of the ISRM, reliability is the extent to which an in-service SOI performs without countable failures during (1) the READY and RUNNING modes of an in-service span, and (2) without countable data failures during any IN-SERVICE or REMOVED mode.'

The standard's reliability measurement elements are: failure occurrence (variable), failure count (variable), time of failure (variable), time between software failures (formula), failure intensity (formula), observed reliability (formula), MTBSF (formula), estimated MTBSF given constant failure intensity (formula), and the Laplace test for failure trend (formula).

## 6.2  Availability

'Availability is the extent to which a system of interest is READY or RUNNING during an in-service span, expressed as the ratio of downtime to uptime.'

'An in-service span is divided into uptime (when a system of interest is READY or RUNNING) and downtime (when a system of interest is INOPERABLE or SUPPORTING.) Downtime is the time spent in the INOPERABLE and SUPPORTING modes. The SUPPORTING mode consists of several kinds of support actions.'

The standard's availability measurement elements are: in-service invariant (formula), uptime (formula), downtime (formula), downtime cycle (formula), downtime cycle count (variable), downtime cycle start (variable), downtime cycle finish (variable), time between support (formula), mean time between support (formula), mean downtime (formula), repair cycle count (formula), time to repair (formula), mean time to repair (formula), in-service availability (formula), inherent availability (formula), achieved availability (formula), operational availability (formula), mean operational availability (formula), unavailability (formula), unplanned downtime budget given availability threshold (formula), and preventative maintenance budget (formula).

## 6.3  Supportability

Supportability characterizes the extent, frequency, and consistency of in-service software support actions, distinct from software maintenance.

'Support actions typically do not result in modification of source code and can be performed with staff, resources, and facilities present in an operational environment. These actions include planned downtime for preventative and adaptive purposes and unplanned downtime for corrective repairs.'

The standard's supportability measurement elements are: support update count (variable), support action cycle count (variable), support cycle start (variable), support cycle finish (variable), support cycle time (formula), support cycle slices and spans (formula), span support cycle time (formula), average support cycle time (formula), and support coefficient of variation (formula).

## 6.4  Recoverability

Software recoverability characterizes the effectiveness of built-in fault tolerance capabilities.

'Recoverability is the ability of a system to mitigate the duration and lossage of impairments. It is the extent to which an SOI can self-correct after operating at a degraded level acceptable to stakeholders or dependent systems. This subclause defines measurements that characterize the effectiveness of an SOI's built-in recovery capabilities. These measurements require the user to interpret them for application-specific impairments and recovery capabilities, and to devise a means to track their effects and duration.'

'An impairment is a change in a resource or operating condition that degrades system of interest capabilities but does not result in an immediate system of interest failure. Impairments may be the result of a system of interest vulnerability or bug, a runtime environment failure, or adverse operating environment conditions.'

The standard's recoverability measurement elements are: impairment occurrence (variable), impairment duration (variable), data lossage (variable), capability lossage (variable), mean time to impairment recovery (formula), mean impairment lossage (formula), recovery success rate (formula), and recovery failure intensity (formula).

## Acknowledgements

Any IEEE standard is the result of collaboration among working group volunteers and IEEE staff, foundations established by earlier working groups, and the influence of practice and theory resulting from decades of application and research. About thirty persons have been directly involved in the discussions and development of the draft of 982-2024. The third edition of the standard would not have been possible nor would it have taken its present form without their contributions. It has been my privilege to work with this team including Pieter Botman, Lynn Robert Carter, Sigrid Eldh, Michael Grottke, Lou Gullo, Jon Hagar, Pratap Lakshman, Phil Laplante, Ann Marie Neufelder, Joanna Olszewska, Annette Reilly, Patricia Roder, and Rob Schaaf. A special thanks to Lance Fiondella, Rajesh Murthy, and Jeffery Voas for their technical insights.

Text indented as a block quote, and figures 1 and 4 are reproduced from the draft standard and used by permission of the IEEE.

## About the Author

Robert V. Binder is a consulting software engineer for RBSC Corporation. He specializes in model-based testing, software process assessment, and dependability optimization. He is an IEEE Life Senior Member. Binder holds an MS EECS from the University of Illinois at Chicago and an MBA from the University of Chicago. Contact him at rvbinder@rbsc.com.

⁷ Matthew Heusser, "Software testing lessons learned from knight capital fiasco," CIO, Aug 14, 2012. Accessed: Aug.24, 2023. [Online]. Available: https://www.cio.com/article/286790/software-testing-lessons-learned-from-knight-capital-fiasco.html

⁸ Brad Templeton, "Cruise Robotaxi hit by on call fire engine. Is it the fire truck's fault?" Forbes, Aug. 18, 2023. Accessed: Aug. 24, 2023. [Online] Available: https://www.forbes.com/sites/bradtempleton/2023/08/18/cruise-robotaxi-hit-by-on-call-fire-engine--is-it-the-fire-trucks-fault/